# Magnetic properties and spin structure of MnO single crystal and powder

X Sun[1], E Feng[2], Y Su[2], K Nemkovski[2], O Petracic[1], T Brückel[1]

[1] Jülich Centre for Neutron Science JCNS and Peter Grünberg Institut PGI, JARA-FIT, Forschungszentrum Jülich GmbH, Leo-Brandt Str. 52425 Jülich, Germany.
[2] Jülich Centre for Neutron Science JCNS at Heinz Maier-Leibnitz Zentrum MLZ, Forschungszentrum Jülich GmbH, Lichtenberg Str. 1, 85748 Garching, Germany.

Email: x.sun@fz-juelich.de

**Abstract**. Zero field cooled (ZFC)/Field Cooled (FC) magnetization curves of a bulk MnO single crystal show a peculiar peak at low temperatures (~ 40 K) similar to the low temperature peak observed in MnO nanoparticles. In order to investigate the origin of this peak, the spin structure of a MnO single crystal has been studied and compared with a single phase powder sample using magnetometry and polarized neutron scattering. Both magnetometry and polarized neutron diffraction results confirm the antiferromagnetic (AF) phase transition at the Néel temperature $T_N$ of 118 K, in both powder and single crystal form. However, the low temperature peak in the ZFC/FC magnetization curves is not observed in single phase MnO powder. To better understand the observed behavior, ac susceptibility measurements have been employed. We conclude that the clear peak in the magnetic signal from the single crystal originates from a small amount of ferrimagnetic (FiM) $Mn_2O_3$ or $Mn_3O_4$ impurities, which is grown at the interfaces between MnO crystal twins.

## 1. Introduction

As one of the first studied materials using neutron scattering, bulk MnO is a textbook AF with a rocksalt crystal structure [1]. A paramagnetic (PM)-to-AF phase transition at the Néel temperature of 118 K is usually observed in bulk MnO [1,2]. However, in recent studies on MnO nanoparticles the expected feature at the Néel temperature is absent in ZFC/FC magnetization curves, but instead a low temperature peak at ~25 K is found [3–6]. This phenomenon in particle systems was explained by either superparamagnetism, superspin glass behavior, diluted AF states, finite size or surface effects [7–11]. However, we observe a similar peak at ~ 40 K in the ZFC/FC curves of a MnO single crystal.

One can speculate that this feature may either originate from a nanosize effect as in the nanoparticles or intrinsically from the frustrated superexchange interactions in MnO. One should note that the AF spin state of MnO is still not well understood despite numerous investigations [1,12–17]. The magnetic moments of $Mn^{2+}$ are reported to align parallel within its (111) planes, while magnetic moments between neighboring planes arrange antiparallel [1,16]. However, the spin direction is reported to be perpendicular [14,16,17] to the [111] direction in contrast to common text book knowledge [18]. Moreover, a similar low temperature anomaly has already been observed in powder [19] and polycrystalline MnO [15]. In order to shed light on this low temperature peak in the magnetometry, we studied the magnetic behavior on a MnO single crystal using dc magnetometry, ac susceptometry and polarized neutron scattering. The results are compared with annealed MnO powder.

For the MnO powder a specific annealing procedure (see below) guaranteed single phase MnO with no $Mn_2O_3$ and $Mn_3O_4$ impurities.

## 2. Experimental

The commercially available MnO single crystal was obtained from Surface Net GmbH with dimensions 5 x 5 x 2 $mm^3$. From a pre-characterization using a Laue diffraction camera we observe two to three crystal twins. Further structural characterizations using X-ray diffractometers confirm the fcc crystal structure of MnO at room temperature. Moreover, no reflection from other phases is observed. The MnO powder (Alfa Aesar GmbH, 99.99%) was heated at 920°C for 150 hours in vacuum to remove nano-sized particles and potentially existing impurity phases $Mn_3O_4$ and $Mn_2O_3$. Subsequently, the powder was heated to 1000°C for several hours to avoid oxidation after cooling down to room temperature [14]. Magnetometry and ac susceptibility measurements have been performed using a SQUID (Superconducting Quantum Interference Device) magnetometer and a PPMS (Physical Property Measurement System) from Quantum Design. The ac excitation amplitude was 0.08 mT. Polarized neutron scattering was performed at the Diffuse Neutron Scattering (DNS) instrument at the Heinz Maier-Leibnitz Zentrum (MLZ) [20,21] with λ = 3.3 Å for powder and 4.2 Å for the single crystal.

## 3. Results and discussions

ZFC/FC magnetization curves of the MnO single crystal are shown in Fig. 1. Besides the expected AF-to-PM phase transition at $T_N$ = 118 K, an additional peak at $T_P$ (peak temperature) ≈ 40 K is observed. A similar peak was observed previously in both MnO nanoparticles [3–6] and powder [19] and was explained by superparamagnetism or spin-glass behavior. However, the ZFC/FC curves on the MnO single crystal in Fig. 1 rather show a FM-like behavior, i.e. a steep drop above $T_p$ and an overall shape of a typical FM order parameter. One should note that X-ray diffraction and X-ray fluorescence analysis on this sample does not show any impurity.

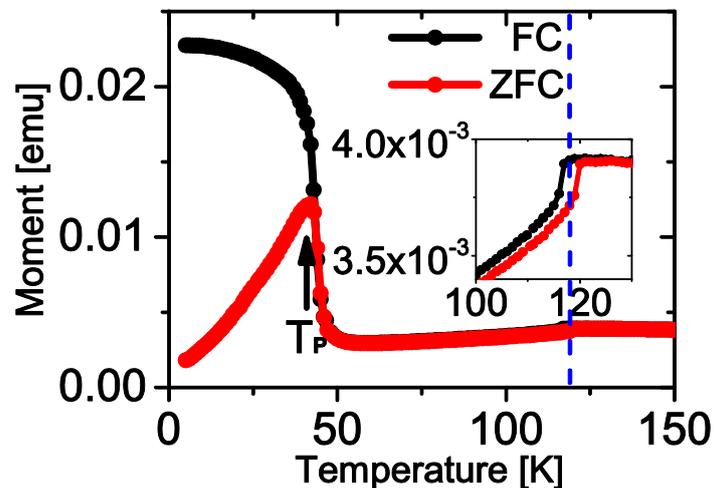

Figure 1: ZFC/FC curves of the MnO single crystal measured at 100 mT. The inset shows the enlarged view near the Néel temperature of MnO at 118 K.

In order to study the origin of this low temperature peak at 40 K in the MnO single crystal, polarized neutron diffraction is performed (Fig. 2). The orientation of the single crystal is fixed on the sample holder so that the $(1\bar{1}0)$ axis is perpendicular to the scattering plane. According to ref. [22,23], only magnetic contributions with spin incoherent background are measured in the spin-flip (sf) channel and the full nuclear coherent contribution with a little bit of incoherent background are measured in the non-spin-flip (nsf) channel, when the polarization ($\vec{P}$) of the neutrons is parallel to the scattering vector ($\vec{Q}$). After corrections of the background, flipping ratio and detector efficiency,

Fig. 2(a) shows the magnetic contribution measured at 4 K. Fig. 2(b) shows the temperature dependence of the magnetic $\left(\frac{\bar{1}}{2} \frac{\bar{1}}{2} \frac{\bar{1}}{2}\right)$ Bragg peak. The AF order parameter of MnO vanishes as expected at $T_N$ = 118 K. However, no magnetic phase transition at $T_P \approx$ 40 K is found. The temperature dependence of the magnetic $\left(\frac{\bar{1}}{2} \frac{\bar{1}}{2} \frac{\bar{3}}{2}\right)$ Bragg peak shows a similar behavior like the magnetic $\left(\frac{\bar{1}}{2} \frac{\bar{1}}{2} \frac{\bar{1}}{2}\right)$ Bragg peak (data not shown). However, at the position of the ($\bar{1}$ $\bar{1}$ $\bar{1}$) nuclear peak of MnO (Q = 2.5 Å$^{-1}$) a small feature is observed at about 40 K before background, flipping ratio and detector efficiency corrections (Fig. 2(c)). This ($\bar{1}$ $\bar{1}$ $\bar{1}$) nuclear peak is located near the expected Q-value of the (2 0 2) and (2 1 1) Bragg peaks of ferrimagnetic (FiM) $Mn_3O_4$. Taking the broadening of the Bragg peak into account, this might be due to the FiM-to-PM transition of $Mn_3O_4$ (Curie temperature $\approx$ 42 K) [24,25]. One should note that for a single crystal a higher oxidation state ($Mn_2O_3$) is not expected. The drop observed at 118 K in Fig. 2(c) is very likely due to the imperfect separation from the nuclear component. Another explanation of this drop is the extinction effect because of the structural distortion of MnO below the Néel temperature.

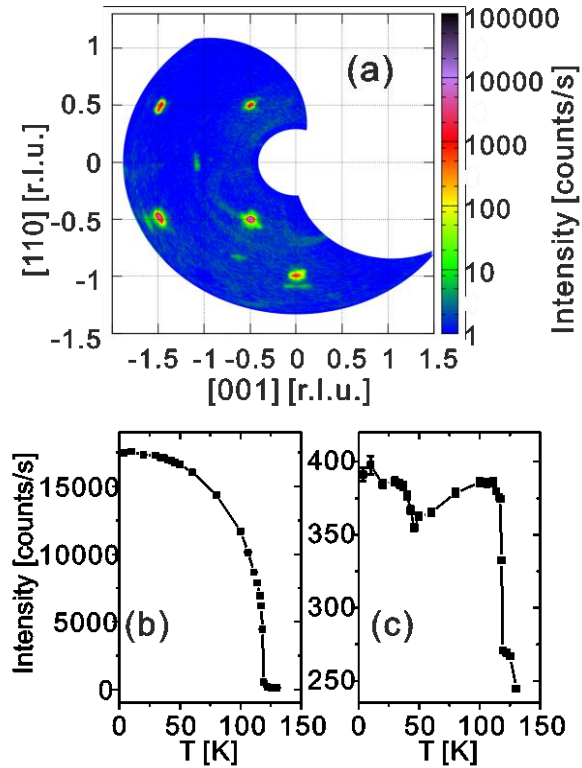

Figure 2: Polarized neutron diffraction of (a) the MnO single crystal (sf channel) measured at 4 K. (b) The intensity of the magnetic $\left(\frac{\bar{1}}{2} \frac{\bar{1}}{2} \frac{\bar{1}}{2}\right)$ Bragg peak of MnO single crystal as function of temperature. (c) The intensity of the sf-channel at the position of ($\bar{1}$ $\bar{1}$ $\bar{1}$) Bragg peak of the MnO single crystal as function of temperature.

Further measurements have been performed on the MnO single crystal to shed light on the behavior of the low temperature peak in the ZFC/FC measurements. AC-susceptibility results show two peaks in the χ'(T) curve (Fig. 3). Both peaks, the one at 40 K and the second at 118 K have no frequency dependence, which is expected for FM/FiM or AF systems [2]. The peak at 118 K indicates the phase transition of bulk MnO from the AF to the PM state. The peak at 40 K very likely marks the PM to FiM transition of $Mn_3O_4$ since the temperature value matches the Curie temperature of 42 K.

However, this feature at 40 K is not observed away from the ($\bar{1}$ $\bar{1}$ $\bar{1}$) peak of MnO. Therefore it is possibly due to intergrowth of $Mn_3O_4$ inside MnO at the interfaces between MnO crystal twins.

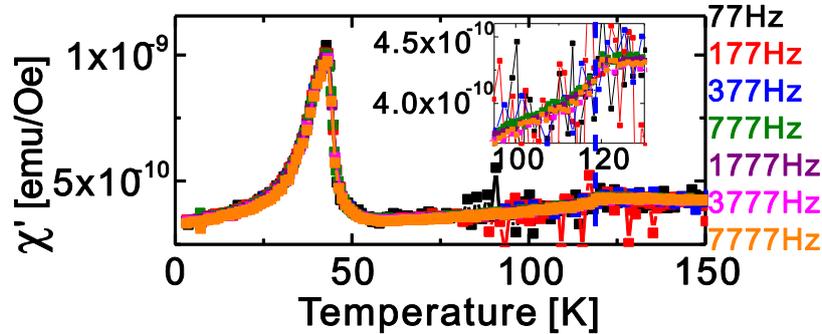

Figure 3: AC-susceptibility measurements of MnO single crystal. The inset is an enlarged view near the Néel temperature of MnO.

From the above measurements we conclude that the MnO single crystal contains $Mn_3O_4$ impurities. We assume that the imperfect crystal has grain boundaries and/or dislocations at which oxidation toward $Mn_3O_4$ occurred. These impurities are invisible in X-ray scattering due to a negligible amount of the $Mn_3O_4$. Interestingly, magnetometry is able to reveal these impurities. This is due to the fact that a FM or FiM subsystem has a much larger magnetic moment compared to a perfectly ordered AF phase, particularly at relatively small applied magnetic fields as in our case. The amount of $Mn_3O_4$ impurities is estimated to be about 0.034 mg assuming the 40 K peak in the ZFC curve originates completely from $Mn_3O_4$. This is about 0.3% of the sample mass.

To further investigate whether the low temperature peak originates from a higher oxidation state or from intrinsical magnetic behavior of MnO alone, a single phase MnO powder sample (after the annealing procedure) was compared with the single crystal results.

Fig. 4(a) shows the ZFC/FC magnetization curves obtained on this MnO powder. As expected only a feature at the Néel temperature of MnO (118 K) is observed. Fig. 4(b) displays the AC-susceptibility measurements of MnO powder. As in the ZFC/FC curves, only one peak indicating the Néel temperature of MnO is found. In addition, this peak is frequency independent as much as one can estimate from the data. The relatively large noise is due to the extremely small signal as expected from a perfect AF at zero applied dc field and hence perfectly compensated AF sublattices.

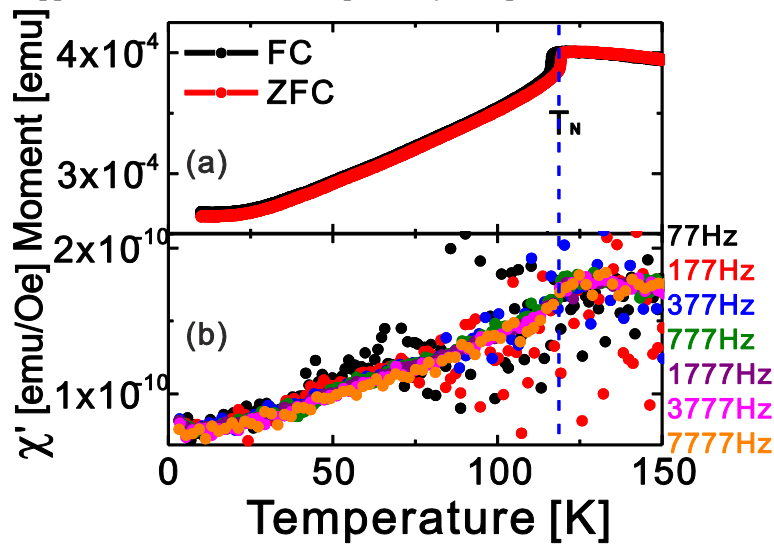

Figure 4: (a) ZFC/FC magnetization curves of MnO powder measured at 100 mT. (b) AC-susceptibility data, χ' on MnO powder performed at various frequencies. The Néel temperature $T_N$ of MnO is marked by the dashed line.

Polarized neutron diffraction was also performed on MnO powder (Fig. 5). Using xyz-polarization analysis [22,23], we obtained separated scattering intensities which are shown in Fig. 5(a). The magnetic $\left(\frac{1}{2} \frac{1}{2} \frac{1}{2}\right)$ and $\left(\frac{3}{2} \frac{1}{2} \frac{1}{2}\right)$ Bragg peaks can be found at Q = 1.25 Å$^{-1}$ and Q = 2.3 Å$^{-1}$, respectively. The nuclear (111) peak is located at Q = 2.5 Å$^{-1}$. The AF-to-PM phase transition of MnO is found at $T_N \approx 120$ K from the magnetic $\left(\frac{1}{2} \frac{1}{2} \frac{1}{2}\right)$ (Fig. 3(b)) Bragg peak. In addition, no feature near 40 K is observed in both magnetic Bragg peaks which agrees with the single crystal results.

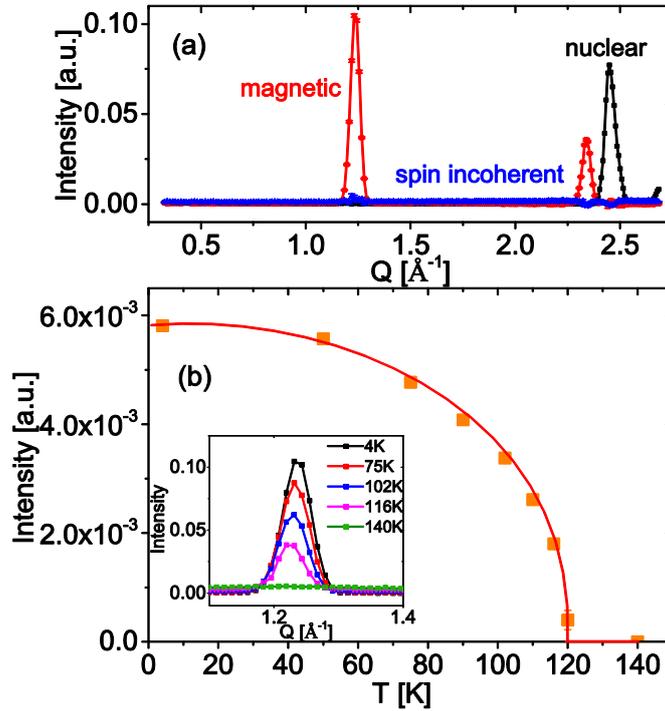

Figure 5: Polarized neutron diffraction of (a) the MnO powder measured at 4 K. (b) The intensity of the magnetic $\left(\frac{1}{2} \frac{1}{2} \frac{1}{2}\right)$ Bragg peak as function of temperature (inset shows the $\left(\frac{1}{2} \frac{1}{2} \frac{1}{2}\right)$ peak at various temperatures). The red line is a guide to the eye.

## 4. Conclusion

Polarized neutron scattering, magnetometry and ac susceptometry results of the MnO single crystal show an anomaly at about 40 K additional to the well-known AF to PM phase transition of MnO at 120 K. This feature at 40 K shows no frequency dependence in the ac susceptibility results, which is different from the spin-glass behavior reported in a previous study [19]. We conclude that the MnO single crystal contains ferrimagnetic $Mn_3O_4$ impurities. Interestingly, the $Mn_3O_4$ impurity is only observed at the position of the $(\bar{1}\,\bar{1}\,\bar{1})$ Bragg peak in polarized neutron diffraction. This indicates that the higher oxidation phase $Mn_3O_4$ crystallizes very likely at the interfaces between crystal twins. These impurities cannot be observed by X-ray scattering alone due to a negligible amount of $Mn_3O_4$. Magnetometry and susceptometry can reveal these impurities since a ferro- or ferrimagnetic subsystem has a much larger magnetic moment compared to a perfectly ordered antiferromagnetic subsystem. Moreover, we exclude that the low temperature peak in magnetometry is due to an intrinsic magnetic behavior of MnO since the single phase MnO powder does not show this feature but only regular antiferromagnetic behavior.

**Acknowledgments**
We would like to thank Hailey Williamson, Pankaj Thakuria, Thomas Müller, Paul Hering, Karen Friese and Jörg Perßon for their help with the structural characterizations of the single crystal and the annealing procedure of the powder sample. We would also like to thank Berthold Schmitz for the technical supports with the magnetometry measurements. We thank ZEA-3: Analytik in Forschungszentrum Jülich for the X-ray fluorescence analysis of the MnO single crystal.

**Supporting information**

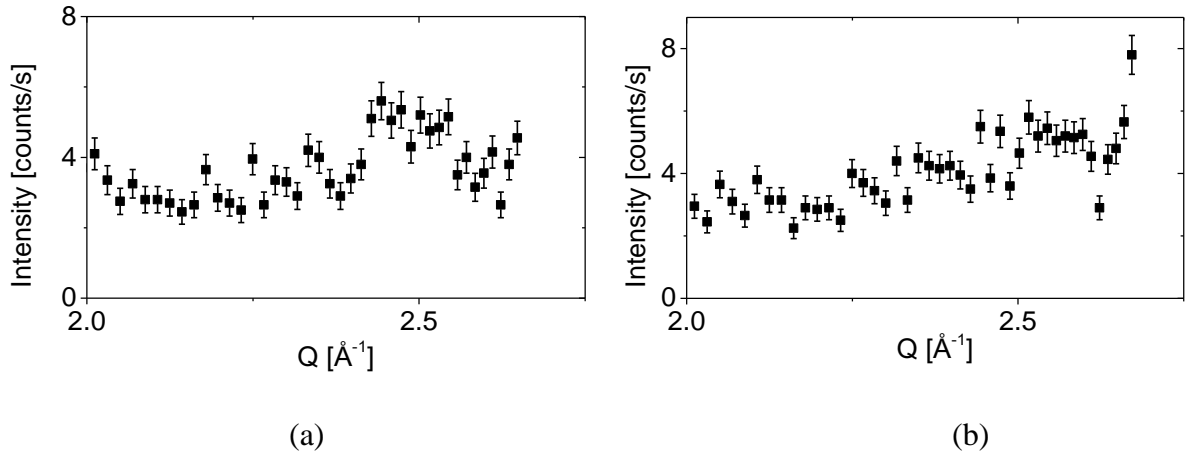

(a)                                                (b)

Fig. 1: Q-dependence measured at the angle of (a) magnetic $(\frac{1}{2}\frac{1}{2}\frac{1}{2})$ peak (71°) and (b) 86° at 4 K. No obvious peak can be observed within the error of the background at $Q = 2.5$ Å$^{-1}$.

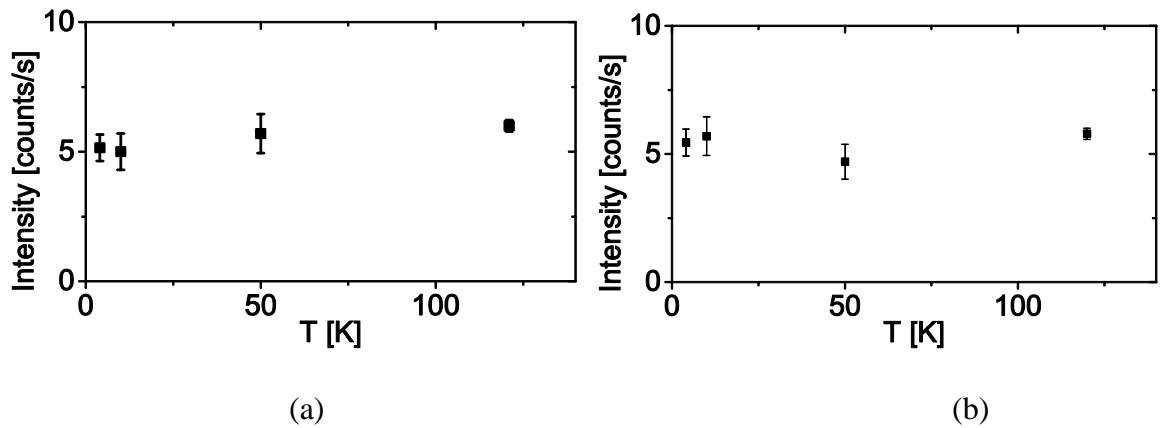

(a)                                                (b)

Fig. 2: Temperature dependence of the intensity at $Q = 2.5$ Å$^{-1}$ away from the $(\bar{1}\bar{1}\bar{1})$ nuclear peak of MnO at (a) 71° and (b) 86°.